\documentclass[twocolumn,aps,prl,showpacs,superscriptaddress,amsfonts,amssymb,amsmath]{revtex4}
\usepackage{graphicx}
\def\s{\sigma}

\def\V={{{\bf\rm{V}}}}

\def\L{\Lambda}

\def\beq{\begin{equation}}
\def\eeq{\end{equation}}
\def\bea{\begin{eqnarray}}
\def\eea{\end{eqnarray}}
\def\ba{\begin{array}}
\def\ea{\end{array}}
\def\no{\nonumber}

\def\lt{\left}
\def\rt{\right}

\begin{document}

\title{Exact ground state and elementary excitations of a topological spin chain}
\author{Yi Qiao}
\affiliation{Institute of Modern Physics, Northwest University,
Xi¡¯an 710127, China}
\affiliation{Beijing National Laboratory for Condensed Matter
Physics, Institute of Physics, Chinese Academy of Sciences, Beijing
100190, China}
\author{Pei Sun}
\affiliation{Institute of Modern Physics, Northwest University,
Xi¡¯an 710127, China}
\author{Junpeng Cao}
\affiliation{Beijing National Laboratory for Condensed Matter
Physics, Institute of Physics, Chinese Academy of Sciences, Beijing
100190, China}
\affiliation{School of Physical Sciences, University of Chinese Academy of
Sciences, Beijing, China}
\affiliation{Songshan Lake Materials Laboratory, Dongguan, Guangdong 523808, China}
\affiliation{Peng Huanwu Center for Fundamental Theory, Xian 710127, China}
\author{Wen-Li Yang}
\thanks{wlyang@nwu.edu.cn}
\affiliation{Institute of Modern Physics, Northwest University,
Xi¡¯an 710127, China}
\affiliation{Peng Huanwu Center for Fundamental Theory, Xian 710127, China}
\affiliation{Shaanxi Key Laboratory for Theoretical Physics Frontiers,  Xian 710127, China}
\author{Kangjie Shi}
\affiliation{Institute of Modern Physics, Northwest University,
Xi¡¯an 710127, China}
\author{Yupeng Wang}
\thanks{yupeng@iphy.ac.cn}
\affiliation{Beijing National Laboratory for Condensed Matter
Physics, Institute of Physics, Chinese Academy of Sciences, Beijing
100190, China}
\affiliation{The Yangtze River Delta Physics Research Center, Liyang, Jiangsu, China}

\begin{abstract}
A novel Bethe Ansatz scheme is proposed to calculate physical properties of quantum integrable systems without $U(1)$ symmetry. As an example, the anti-periodic XXZ spin chain, a typical correlated many-body system embedded in a topological manifold, is examined. Conserved  ``momentum" and ``charge" operators are constructed despite the absence of translational invariance and $U(1)$ symmetry. The ground state energy and elementary excitations are derived exactly. It is found that two intrinsic fractional (one half) zero modes accounting for the double degeneracy exist in the eigenstates. The elementary excitations show quite a different picture from that of a periodic chain. This method can be applied to other quantum integrable models either with or without $U(1)$ symmetry.
\end{abstract}



\maketitle
Exact solution of  quantum integrable systems without $U(1)$ symmetry is an important issue in modern mathematical physics. It is related to a number of important problems such as exact quantization in the string theory\cite{nek,hua,vaf} and topological states of matters \cite{m2,m3} in correlated condensed matter systems, since exact solutions often provide useful benchmarks to understand relevant physical phenomena. A formidable problem to solve such kind of integrable models is the absence of $U(1)$ symmetry, which makes us frustrated to work in a traditional particle-hole representation. Though some methods \cite{cao03,nep02,Bas1,gal07,sk1,sk2,Bel13} have been developed to approach this problem, including the off-diagonal Bethe Ansatz \cite{cysw,Zhan14} proposed by some of the present authors, with which the formal solutions of the eigenvalues can be expressed in an inhomogeneous $t-Q$ relation, how to construct elementary excitations and to calculate physical properties of these models remain still unclear because of complicated distribution of Bethe roots associated with inhomogeneous Bethe Ansatz equations.

In this letter, we propose a novel Bethe Ansatz scheme to obtain exact quantized spectrum of the topological quantum spin chain model. By constructing an operator identity of the transfer matrix for arbitrary spectral parameter $u$, factorized Bethe Ansatz equations (BAEs) about the roots  of the transfer matrix are derived, which permit us to define quantum winding numbers associated with the roots. The nice pattern of the root distribution in the complex plane allows us to construct the exact ground state and to study what kind of elementary excitations may emerge. A counterpart of momentum operator and a conserved charge in the topological manifold are also defined. The topological momentum operator and the $Z_2$ operators together can be used to classify the eigenstates.

To clarify our procedure clearly, we study the anti-periodic XXZ spin chain \cite{G1,G2} as a concrete example. The model Hamiltonian reads
\begin{eqnarray}
H = -\sum_{n=1}^N
  (\sigma_n^x\sigma_{n+1}^x+\sigma_n^y\sigma_{n+1}^y
  +\cosh\eta \,\sigma_n^z\sigma_{n+1}^z), \label{xxzh}
\end{eqnarray}
with the topological boundary condition
\bea
\sigma^{\alpha}_{1+N}=\sigma^x_1\, \sigma^{\alpha}_{1}\,\sigma^x_1,\quad {\rm  for}\quad \alpha=x,y,z,\label{BC}
\eea
where $\sigma^x,\,\sigma^y,\,\sigma^z$ are the usual Pauli
matrices and $\eta$ is the coupling constant. This nontrivial boundary condition mixes the spin up and spin down states in the Hilbert space and makes the system form a quantum M{\"o}bius strip. The $U(1)$ symmetry is thus broken and a discrete $Z_2$ invariance $[H,
U_\alpha]=0$ is left with $U_\alpha=\prod_{j=1}^N\sigma^\alpha_j$ and $U_\alpha^2=1$. $U_\alpha$ (only two of them are independent) are the generators of the $Z_2$ algebra. In the following text, we put $\eta=i\gamma$ as an imaginary constant. The real $\eta$ case can be studied straightforwardly. The integrability of the model is associated with the well-known six-vertex $R$-matrix
\bea
R_{0,j}(u)&=&\frac{\sinh(u+\eta)+\sinh u}{2\sinh \eta}+\frac{1}{2} (\sigma^x_j \sigma^x_0 +\sigma^y_j \sigma^y_0) \nonumber\\
 &&+ \frac{\sinh(u+\eta)-\sinh u}{2\sinh \eta} \sigma^z_j \sigma^z_0,
\label{r-matrix} \eea
which satisfies the Yang-Baxter equation \cite{yang2,bax1}, where $u$ is the spectral parameter.

Let us introduce the monodromy matrix
\begin{eqnarray}
{\bf T}_0(u)=R_{0,N}(u)\cdots R_{0,1}(u).\label{Mon-1}
\end{eqnarray}
The
transfer matrix ${\bf t}(u)$  is given by \cite{G1}
 \begin{eqnarray}
 {\bf t}(u)=tr_0\lt\{\sigma^x_0\,{\bf T}_0(u)\rt\},\label{trans-per}
 \end{eqnarray} where $tr_0$ denotes trace over the
``auxiliary space" $0$. From Eqs.(\ref{r-matrix})-(\ref{trans-per}) we know that ${\bf t}(u+i\pi)=(-1)^{N-1}{\bf t}(u)$ and
${\bf t}(u)$ as a function of $u$, is an operator-valued trigonometric polynomial of degree $N-1$. The Hamiltonian described by (\ref{xxzh}) and (\ref{BC}) is given by
\begin{eqnarray}
H=-2\sinh\eta \,\frac{\partial \ln{\bf t}(u)}{\partial
u}|_{u=0}+N\cosh\eta.\label{ham}
\end{eqnarray}

The commutativity of the transfer matrices with different spectral parameters ensured by the Yang-Baxter equation
implies that their eigenstates are $u$-independent. Given an eigenstate $|\Psi\rangle$ with
eigenvalue $\Lambda(u)$,
\bea
{\bf t}(u)|\Psi\rangle=\Lambda(u)|\Psi\rangle,\no
\eea
we have
$\L(u+i\pi)=(-1)^{N-1}\L(u)$. $\Lambda(u)$  as a function of $u$, is a trigonometric polynomial of degree $N-1$ and can be expressed in terms of its $N-1$ zero roots
$\{z_j-\eta/2|j=1,\cdots,N-1\}$ and an overall  coefficient
$\Lambda_0$ as follows
\bea
\Lambda(u)=\Lambda_0\,\prod_{j=1}^{N-1}\,\sinh (u-z_j+\frac\eta2).\label{Zero-points}
\eea
The corresponding eigenvalue
of the Hamiltonian given by (\ref{ham}) can be expressed  as
\bea
E=2\sinh\eta\,\sum_{j=1}^{N-1}\coth (z_j-\frac\eta2) +N\cosh\eta.
\eea

{\sl Bethe Ansatz:}
The key point of the present Bethe Ansatz scheme is to construct an operator identity for the transfer matrix. We note that  $R_{1,2}(-\eta)=-2P^{(-)}_{1,2}$ and
$P^{(\pm)}_{1,2}=(1\pm P_{1,2})/2$, where $P_{1,2}^{(\pm)}$ and $P_{1,2}$ are the projection operators and permutation operator, respectively. With the fusion techniques \cite{fus,resh}
\begin{eqnarray}
&&{\bf t}(u){\bf t}(u-\eta)
=tr_{1,2}\{P^{(-)}_{1,2}\s^x_1\s^x_2{\bf T}_2(u)
{\bf T}_1(u-\eta)P^{(-)}_{1,2}\}\no\\[4pt]
&&\quad +tr_{1,2}\{P^{(+)}_{1,2}\s^x_1\s^x_2
{\bf T}_2(u){\bf T}_1(u-\eta)P^{(+)}_{1,2}\},
\end{eqnarray}
we have the following $t-W$ relation
\begin{eqnarray}
{\bf t}(u){\bf t}(u-\eta)=-a(u)d(u-\eta)\times{\bf id}+d(u){\bf W}(u),\label{tw}
\end{eqnarray}
where
\begin{eqnarray}
a(u)=d(u+\eta)=\frac{\sinh^N(u+\eta)}{\sinh^N\eta},
\end{eqnarray}
${\bf W}(u)$ is an operator-valued
degree $N$ trigonometric polynomial of $u$ with
$[{\bf W}(u), {\bf t}(v)]=0$;
and ${\bf id}$ is the identity operator in the Hilbert space. Acting (\ref{tw}) on an eigenstate $|\Psi\rangle$ we have
\begin{eqnarray}
\Lambda(u)\Lambda(u-\eta)=-a(u)d(u-\eta)+d(u)W(u),\label{t-w}
\end{eqnarray}
where $W(u)$ is the eigenvalue of ${\bf W}(u)$. Let
\begin{eqnarray}
W(u)=W_0\sinh^{-N}\eta\,\prod_{l=1}^N\sinh(u-w_l),
\end{eqnarray}
with $W_0$ a constant (depending on the roots). An important fact is that (\ref{t-w}) is a degree $2N$ polynomial equation and thus gives $2N+1$ independent equations for the coefficients, which determines the $N-1$ $z_j$ roots, $N$ $w_l$ roots and the two constants $\Lambda_0$ and $W_0$ completely.  Since $\Lambda(u)$ is a degree $N-1$ trigonometric polynomial of $u$, the leading terms in the right hand side of (\ref{t-w}) must be zero. Therefore,
$W_0e^{\pm\sum_{l=1}^N w_l}=1$,
or $W_0^2=1$ and  $\sum_{l=1}^N w_l=0{~~} mod(i\pi)$.
Putting $u=z_j-\eta/2$ in (\ref{t-w}) we obtain
\begin{eqnarray}
\hspace{-0.8truecm}&&\hspace{-0.8truecm}\sinh^N(z_j-\frac{3\eta}2)\,\sinh^N(z_j+\frac\eta2)\nonumber\\
&&\quad=W_0\,\sinh^N (z_j-\frac\eta2)\,\prod_{l=1}^N\sinh(z_j-w_l-\frac\eta2).\label{BA1}
\end{eqnarray}
Putting $u=w_l$ in (\ref{t-w}) we obtain
\begin{eqnarray}
\hspace{-0.8truecm}&&\hspace{-0.8truecm}\Lambda_0^2\prod_{j=1}^{N-1}\sinh(w_l-z_j+\frac\eta2)\sinh(w_l-z_j-\frac\eta2)\nonumber\\
&&=-\sinh^{-2N}\eta \sinh^N(w_l+\eta)\sinh^N(w_l-\eta).
\end{eqnarray}
The coefficient $\Lambda_0$ can be determined by putting $u=0$ in (\ref{t-w}) as
\begin{eqnarray}
\Lambda_0^2\prod_{j=1}^{N-1}\sinh(z_j+\frac\eta2)\sinh(z_j-\frac\eta2)=(-1)^{N-1}.
\end{eqnarray}

From the intrinsic properties of the $R$-matrix, for imaginary $\eta$ we have
\begin{eqnarray}
{\bf t}^\dagger (u)&=& (-1)^{N-1}{\bf t}(u^*-\eta), \no \\
\Lambda(u)&=& (-1)^{N-1} \Lambda^*(u^*-\eta).
\end{eqnarray}
The above relation implies that if $z_j$ is a root, $z_j^*$ must also be a root! Therefore, $z_j$ can be classified into 3 sets:
(1) real $z_j$; (2) ${\rm Im} z_l=-i\pi/2$ (this is because its conjugate shifted by $i\pi$ becomes itself); (3) complex conjugate pairs. Similarly, we have $W^*(u^*)=(-1)^NW(u)$, indicating that if $w_l$ is a root of $W(u)$, $w_l^*$ must also be a root! In fact, both the exact numerical solutions for finite $N$ and analytic analysis in the thermodynamic limit (as shown below) indicate that the imaginary parts of a $z$-root conjugate pair are around $\pm n\eta/2$ with $n\geq 2$ a positive integer. For $n=2$, the $z$-root conjugate pair is accompanied by a  $w$-root conjugate pair with imaginary parts around $\pm 3\eta/2$. For $n>2$, the $z$-root conjugate pair is accompanied by a $w$-root 4-string with imaginary parts $\pm(n-1)\eta/2$ and $\pm(n+1)\eta/2$. Such a simple pattern of the roots is quite similar to the string structure appeared in the conventional Bethe Ansatz solvable models \cite{takahashi} and allows us to calculate physical properties exactly in the thermodynamic limit. The exact diagonalization of the transfer matrix up to $N=12$ was performed numerically and all the roots solved indeed exactly coincide with those by solving the BAEs (13)-(15). The numerical results for $N=4$ and $\eta=0.6i$ are shown in Table I.
\begin{table}[!h]
\centering
\caption{$z$ roots calculated via exact numerical diagonalization of the transfer matrix  for $N=4$ and $\eta=0.6i$. Each set of solutions is doubly degenerate due to the $Z_2$ symmetry.}
\begin{tabular}{ccc}
\hline\hline $ z_1 $ & $ z_2 $ & $ z_3 $\\ \hline
$-0.2890$ & $0.0000$ & $0.2890$  \\
 $-1.4697$ & $-0.0531$ & $0.2266$ \\
 $-0.2266$ & $0.0531$ & $1.4697$ \\
 $-0.1490$ & $0.0000-1.5708i$ & $0.1490$ \\
 $-0.7908-1.5708i$ & $0.0000$ & $0.7908-1.5708i$ \\
 $-0.1652$ & $0.1384-0.6102i$ & $0.1384+0.6102i$ \\
 $-0.1384-0.6102i$ & $-0.1384+0.6102i$ & $0.1652$ \\
 $0.0000-1.5708i$ & $0.0000-0.6238i$ & $0.0000+0.6238i$\\
 \hline\hline
 \end{tabular}
\end{table}

{\sl Conserved quantities:} Due to the topological boundary, the model possesses neither translational invariance nor $U(1)$ symmetry. Nevertheless we find that
\begin{eqnarray}
{\bf t}(0)=\sigma_1^xP_{1,N}P_{1,N-1}\cdots P_{1,2},
\end{eqnarray}
is a conserved quantity and represents the shift operator in the topological manifold. A corresponding  ``momentum" operator can thus be defined as ${\bf P}_q=-i\ln {\bf t}(0)$. From the definition of the transfer matrix we have ${\bf t}^{2N}(0)=1$, indicating that the eigenvalues of ${\bf P}_q$ take values of
\begin{eqnarray}
k=\frac{\pi l}{N}{~~}mod\,\{\pi\},
\end{eqnarray}
with  $l=\{-N, -N+1,\cdots, N-1\}$.
The topological momentum is related to the $z$-roots as
\begin{eqnarray}
k=-\frac i2\sum_{j=1}^{N-1}\ln\frac{\sinh(z_j+\frac\eta2)}{\sinh(z_j-\frac\eta2)}+(1-(-1)^{N-1})\frac\pi4.\label{k}
\end{eqnarray}
Therefore, the $z$-roots play the roles of  quasi momenta as the Bethe roots do in the conventional Bethe Ansatz.

Similarly, we have the following conserved charge operator
\begin{eqnarray}
{\bf M}_q&=&\frac12({\bf l}_q^++{\bf l}_q^-)\no\\
&=&\frac14e^{-\frac{(N-1)\eta}2}\lim_{u\to\infty}(2\sinh\eta e^{-u})^{N-1}{\bf t}(u),
\end{eqnarray}
where
\begin{eqnarray}
{\bf l}_q^\pm=\frac12\sum_{j=1}^N e^{\mp\frac\eta2\sum_{k=j+1}^N\sigma_k^z}\sigma_j^\pm e^{\pm\frac\eta2\sum_{k=1}^{j-1}\sigma_k^z},
\end{eqnarray}
are two generators of the quantum group \cite{qg} associated with the model. The corresponding eigenvalues of the operator ${\bf M}_q$ is given by
\begin{eqnarray}
M_q&=&\frac14\sinh^{N-1}\eta\,\Lambda_0\,e^{-\sum_{k=1}^{N-1}z_k}.
\end{eqnarray}
Only when $\eta\to 0$, the model tends to an isotropic spin chain and the $U(1)$ symmetry recovers with ${\bf M}_q=\sum_{j=1}^N \sigma_j^x/2$, which is just the $U(1)$ charge. We note that ${\bf M}_q$ is not an $U(1)$ charge for generic $\eta$.
\par
{\sl Ground state:}
For the ground state, all roots $z_j$ and $w_l$ take real values around zero symmetrically. Taking the logarithms of (\ref{BA1}) and its complex conjugate we have
\begin{eqnarray}
  2\theta_1(z_j)-\theta_3(z_j)=\frac{4\pi I_j}N-\frac1N\sum_{l=1}^N\theta_1(z_j-w_l),\label{bae1}
\end{eqnarray} and
\begin{eqnarray}
\ln|\Lambda_0\sinh(z_j\hspace{-0.08truecm}-\hspace{-0.08truecm}\frac{3\eta}2)|=
\frac1N\sum_{l=1}^N\ln|\sinh(z_j\hspace{-0.08truecm}-\hspace{-0.08truecm}w_l
\hspace{-0.08truecm}-\hspace{-0.08truecm}\frac\eta2)|,\label{bae2}
\end{eqnarray}
where $I_j$ denote the quantum numbers (integers or half odd integers depending on the parity of $N$) associated with the root $z_j$  and $\theta_n(x)=2\cot^{-1}(\coth x \tan\frac{n\gamma}{2})$. The quantum numbers take values
\begin{eqnarray}
I_j&=&\left\{-\frac {N-2}2,-\frac {N-4}2,\cdots,\frac {N-4}2, \frac {N-2}2\right\}.\no
\end{eqnarray}
In the thermodynamic limit $N\to\infty$, we define the density of $z$-roots and the density of $z$-holes per unit site as $\rho(z)$ and $\rho^h(z)$, the density of $w$-roots as $\sigma(w)$, respectively. Taking the continuum limits of (\ref{bae1}) and (\ref{bae2}) we have
\begin{eqnarray}
2a_1(z)-a_3(z)&=&2\rho(z)+2\rho^h(z)-a_1*\sigma(z),\\
b_3(z)&=& b_1*\sigma(z),
\end{eqnarray}
where $a_n(z)=\theta_n'(z)/(2\pi)$, $b_n(z)=\ln'|\sinh(z-n\eta/2)|/\pi$ and $*$ indicates convolution. With Fourier transformation we readily have
\begin{eqnarray}
  \rho(z)+\rho^h(z)= \frac{2\cosh(\frac{\pi z}{\pi-\gamma}) \sin(\frac{\pi\gamma}{2\pi-2\gamma})}{(\pi-\gamma)[\cosh(\frac{2\pi z}{\pi-\gamma}) +\cos(\frac{\pi(\pi-2\gamma)}{\pi-\gamma})]}.\no
\end{eqnarray}
$\rho^h(z)$ is non-zero only in the range $|z|>D$ ( $D\to\infty$ in the thermodynamic limit) with $N\int_D^\infty\rho^h(z)dz=1/2$ and $N\int_{-\infty}^{-D}\rho^h(z)dz=1/2$. The existence of the hole density is due to the fact that the total number of roots must be $N-1$ while the dimension of  the Brillouin zone is $N$. However, the hole separates into two halves due to the topological restriction and each half hole locates at one edge of the spectral space. Clearly, the two half-holes contribute two half zero modes (carrying zero energy). Algebraically, we do have a generator of such zero modes. If $|\Psi\rangle$ is a common eigenstate of $H$ and ${\bf P_q}$, then $U_z$ acting on $|\psi\rangle$  generates another degenerate eigenstate because $[H,U_z]=0$ and $[{\bf P_q}, U_z]\neq 0$. The ground state energy density reads
\begin{eqnarray}
e_g&=&-\sin\gamma
 \int
  \frac{\cosh[\frac{(\pi-2\gamma)\tau}{2}]
  \tanh[\frac{(\pi-\gamma)\tau}{2}]} {\sinh(\frac{\pi\tau}{2})}d\tau+\cos\gamma, \no
\end{eqnarray}
which is the same to that of the periodic chain \cite{Yang66-2}.

From the exact numerical diagonalization results we find that besides the ground state there exist several sets of real $z_j$ solutions for small $\gamma$ but their distributions are asymmetric around the origin. Correspondingly, a boundary conjugate pair $\beta\pm m\eta/2$ exists in the set of $w$-roots. A typical set of such solutions is shown in Fig.1.
\begin{figure}[ht]
\centering
\includegraphics[width=7cm]{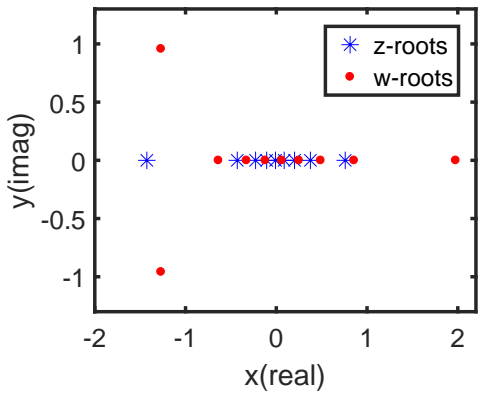}
\caption{Asymmetric real $z$-roots for $N=10$ and $\eta=0.6i$ via exact numerical diagonalization.}
\label{fig-eeh}
\end{figure}In the thermodynamic limit, $\beta$ tends to $\pm\infty$ for keeping the density functions to be convergent; and to ensure the associated energy to be real, $m$ can only take values of odd integers ($\geq3$), coinciding with the numerical results. In this case, the energy is almost degenerate to that of the ground state but the Majorana-like zero modes disappear due to the asymmetric root distribution. Note the double degeneracy still exists since the $w$-pair can locate either on right side or left side of the root distribution.

{\sl Elementary excitations I:} The first kind of elementary excitations is described by a single root locating in the axis ${\rm Im} z=-i\pi/2$ and all the other roots remaining in the real axis. Corresponding to such an excitation, a set of roots derived by exact numerical diagonalization for $N=10$ is shown in Fig.2(a). Let us denote the single complex root as $z=\alpha-i \pi/2$, with $\alpha$ a real number. Accordingly, two $w$-roots form a conjugate pair
$w_\pm=\beta\pm m\eta/2$ with $\beta$ and $m$ two real numbers, and all the other $w$-roots keep real.
\begin{figure}[t]
\begin{center}
\includegraphics[width=4cm]{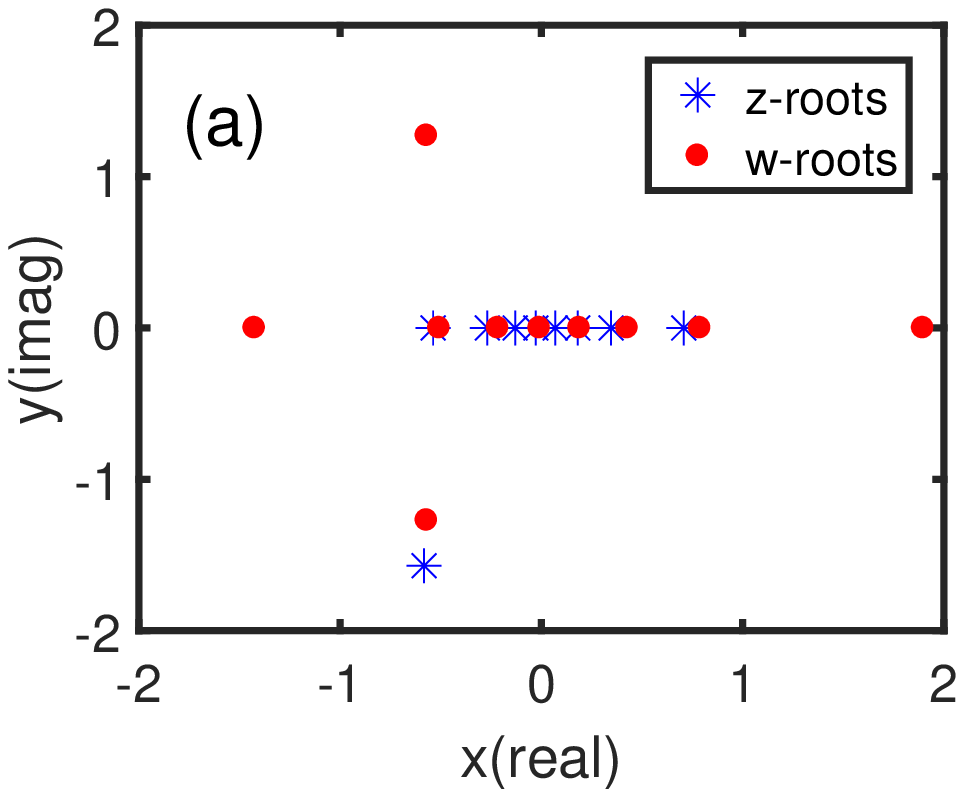}
\includegraphics[width=4cm]{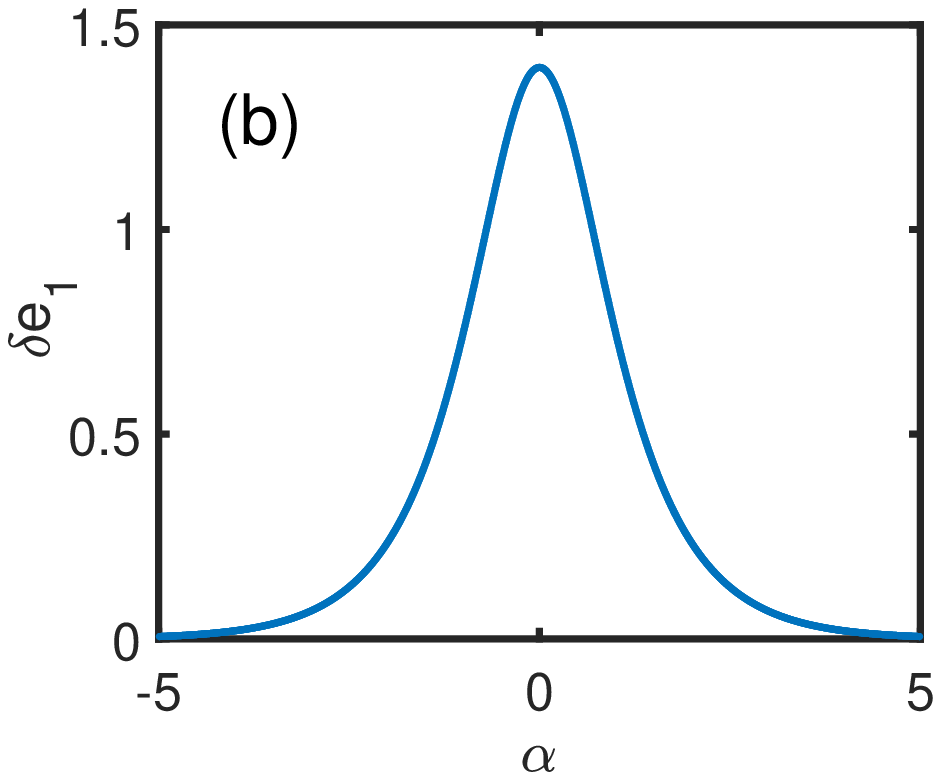}
\caption{(a) A set of zero roots calculated via exact numerical diagonalization for $N=10$ and $\eta=0.6i$.(b) The excitation energy versus $\alpha$ in the thermodynamic limit.}
\end{center}
\label{fig-ee1}
\end{figure}
In the thermodynamic limit, by taking the complex roots into account, we can derive the density $\rho(z)$ for real $z_j$. To ensure the convergence of the density function, the following constraints are needed
\begin{eqnarray}\label{13}
m+1-\frac{\pi}{\gamma}=0,\quad \beta=\alpha.
\end{eqnarray}
The above relations not only fix the relative value between the complex $z$-root and the $w$ conjugate pair but also the imaginary parts of the $w$ conjugate pair. The associated excitation energy reads
\begin{eqnarray}
\delta e_1  &=& \sin\gamma\int \frac{\cos(\tau\alpha)\tanh[\frac{(\pi-\gamma)\tau}{2}]
 \cosh(\frac{\tau\gamma}{2})}{\sinh(\frac{\pi\tau}{2})} d\tau \no \\ && + \frac{2\sin^2\gamma}{\cosh(2\alpha)+\cos\gamma}.
\end{eqnarray}
The momentum associated with $\alpha$ is determined by (\ref{k}). The single parameter $\alpha$ dispersion also indicates some topological confinement of the ``particle" in the ${\rm Im} z=-i\pi/2$ axis and the ``hole" in the real axis as that in the $\eta=i\pi/2$ case, where this kind of excitations is the only possible one\cite{cysw}.

\begin{figure}[ht]
\centering
\includegraphics[width=4cm]{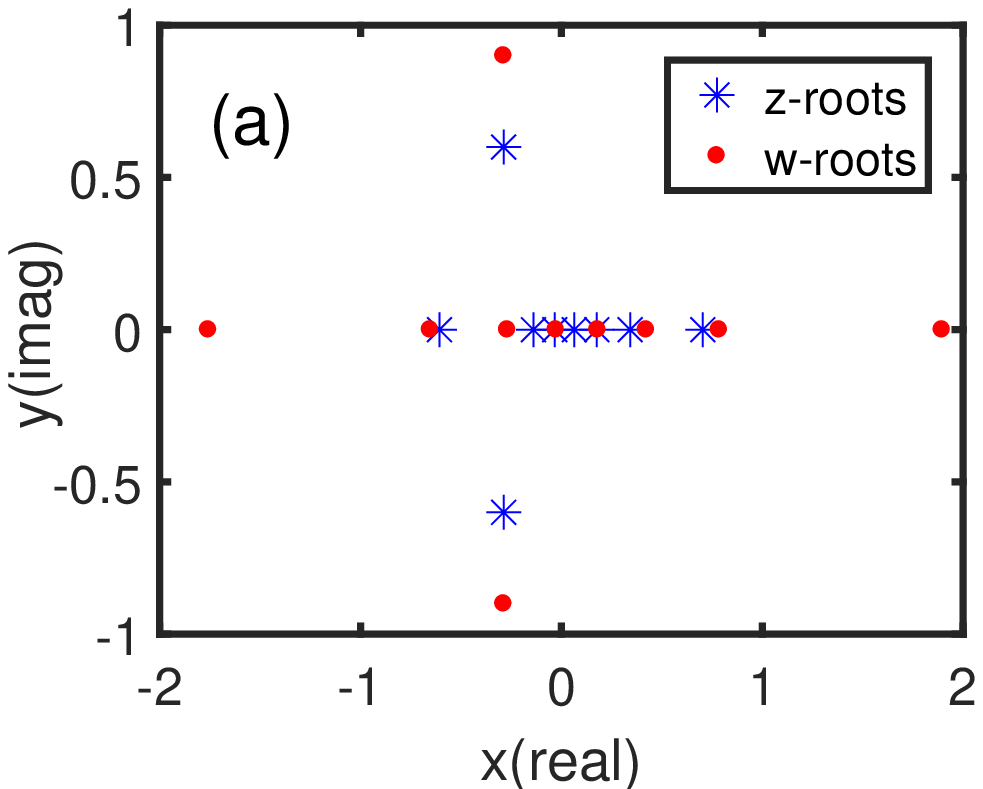}
\includegraphics[width=4cm]{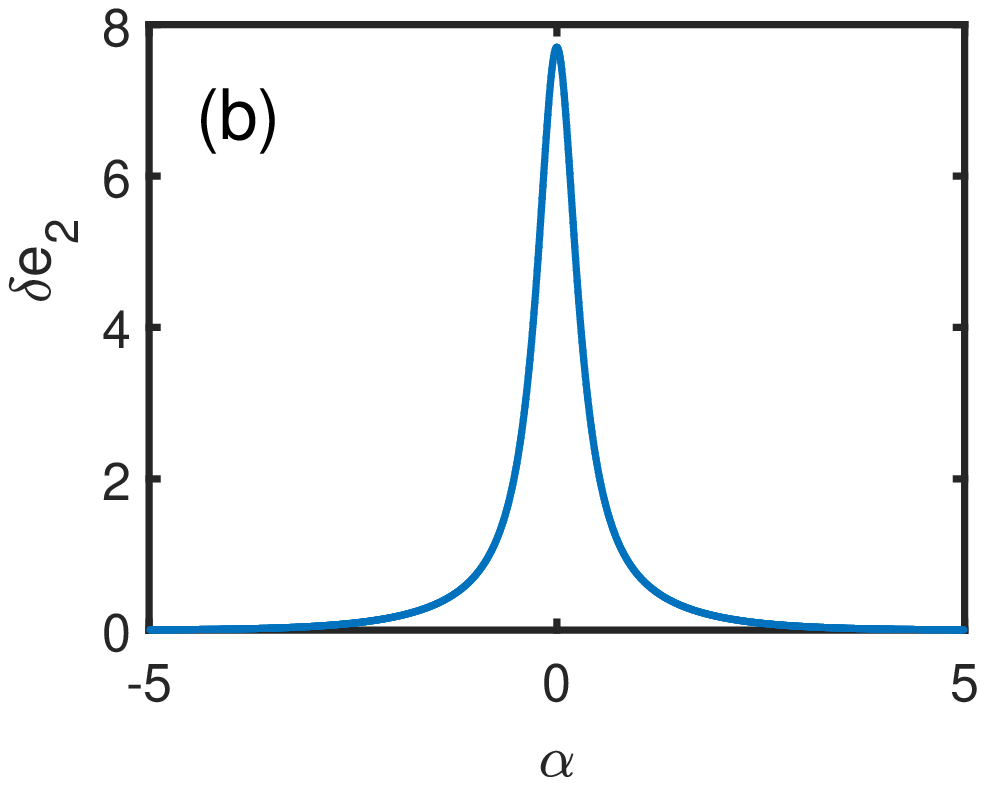}\\
\includegraphics[width=4cm]{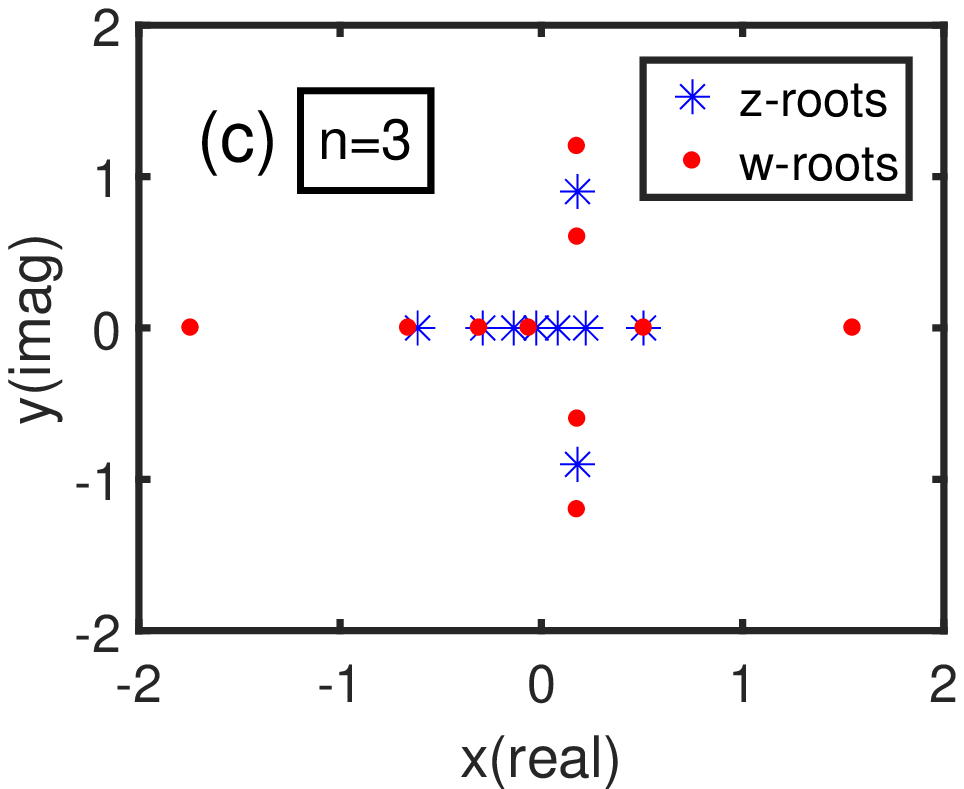}
\includegraphics[width=4cm]{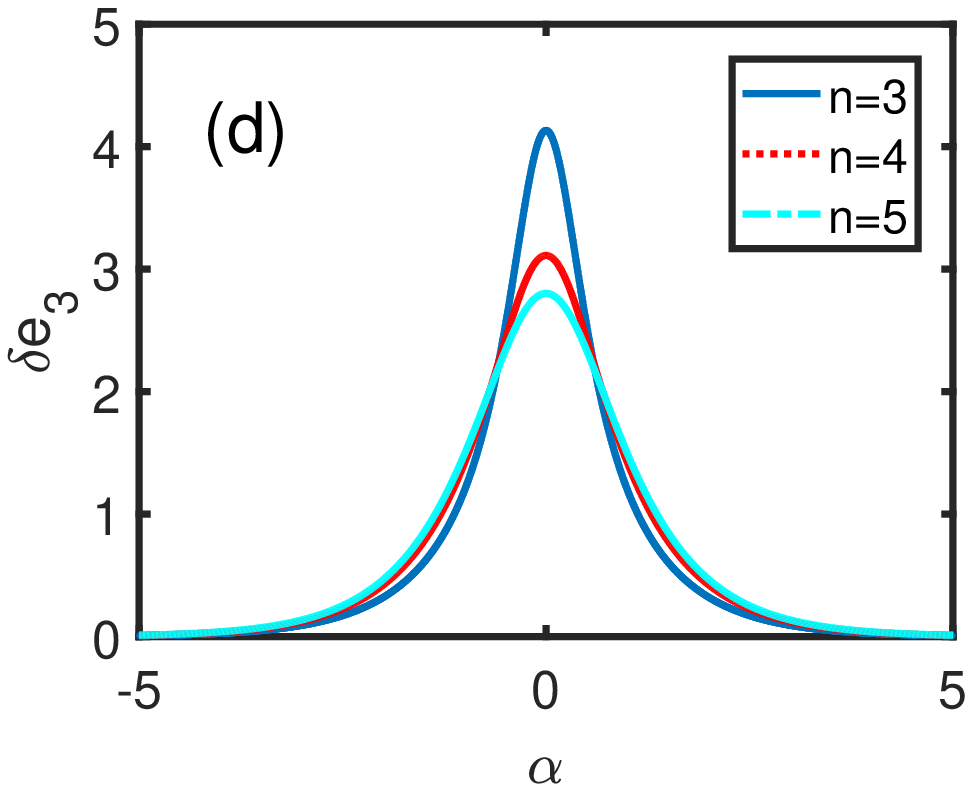}
\caption{(a) A set of zero roots denoting an $n=2$ excitation for $N=10$, $\eta=0.6i$ and (b) the type II excitation energy in the thermodynamic limit. (c) A set of zero roots denoting an $n>2$ excitation for $N=10$, $\eta=0.6i$ and (d) the type III excitation energy in the thermodynamic limit.}
\label{fig-ee2}
\end{figure}

{\sl Elementary excitations II:} When $\eta$ is away from $i\pi/2$, correlation is introduced and conjugate pairs of $z$-roots can exist. Here we consider a single conjugate pair excitation. The simplest conjugate pair is given by $z_\pm\sim \alpha\pm\eta$. The corresponding $w$-set is formed by a $w$ conjugate pair $w_\pm\sim\alpha\pm3\eta/2$ and $N-2$ real $w_l$. Both the positions and the imaginary parts of the conjugate pairs are determined by convergence requirement of the density functions. A corresponding set of roots for $N=10$ is shown in Fig.\ref{fig-ee2}(a).
In the thermodynamic limit, with a similar procedure used in the above text we obtain the energy of this excitation as
\begin{eqnarray}
\delta e_2 &=& \sin\gamma \int \frac{\cos(\tau\alpha)
  \tanh[\frac{(\pi-\gamma)\tau}{2}]
  \cosh[\frac{(\pi-3\gamma)\tau}{2}]}
  {\sinh(\frac{\pi\tau}{2})} d\tau \no \\ &+&
  \frac{4\sin^2\gamma}
  {\cosh(2\alpha)-\cos\gamma}- \frac{2\sin\gamma \sin(3\gamma)}{\cosh(2\alpha)-\cos(3\gamma)}.
\end{eqnarray}

{\sl Elementary excitations III:} General complex-root excitation is described by a conjugate pair $z_\pm\sim\alpha\pm n\eta/2$ with $n\geq3$, and all the other $z$-roots remaining in the real axis. In this case, the corresponding $w$-set is formed by a four-string $\sim\alpha\pm(n+1)\eta/2$, $\alpha\pm(n-1)\eta/2$ and $N-4$ real $w$-roots.
A corresponding set of roots for $N=10$ is shown in Figs.\ref{fig-ee2}(c). In the thermodynamic limit, the excitation energy reads
\begin{eqnarray}
\delta e_3&=&2\sin\gamma\int\frac{\cos(\tau\alpha)\tanh(\frac{\pi-\gamma}{2}\tau)f(\tau)}{\sinh(\frac{\pi\tau}{2})} d\tau \no\\
&&+ \frac{2\sin\gamma\sin[(n-1)\gamma]}{\cosh(2\alpha)-\cos[(n-1)\gamma]}\no\\
&&-\frac{2\sin\gamma\sin[(n+1)\gamma]}{\cosh(2\alpha)-\cos[(n+1)\gamma]},
\end{eqnarray}
where $f(\tau)=\cosh[(1-\delta_{n-1}-\delta_{n+1})\pi\tau/2]\cosh[(\delta_{n-1}-\delta_{n+1})\pi\tau/2]$ and $\delta_m =m\gamma/(2\pi)-\lfloor m\gamma/(2\pi)\rfloor$.

We remark that there is indeed intrinsic difference between elementary excitations in the topological boundary case and those in the periodic boundary case. In the topological case, the dispersion of the excitation energy relies only on a single parameter $\alpha$ (or its corresponding quantum number $I_\alpha/N$) besides the number $n$ ; while in the periodic boundary case, at least two parameters (quasi momenta of two holes in terms of Bethe roots in the real axis) appear in the energy dispersion relation corresponding to two spinons \cite{taka}. Such a phenomenon reveals the topological confinement effect in the elementary excitations of the present model.

In conclusion, a novel Bethe Ansatz scheme is proposed to calculate physical properties of the topological $XXZ$ spin chain by deriving a set of homogeneous BAEs. The exact ground state and elementary excitations are constructed. It is found that fractional zero modes exist in this closed ring and the excitations possess momentum-locked effect. This scheme could be applied to study other quantum integrable models without $U(1)$ symmetry such as open boundary systems with generic off-diagonal boundaries.

The financial supports from National Program for Basic
Research of MOST (Grant Nos. 2016 YFA0300600 and 2016YFA0302104),
National Natural Science Foundation of China (Grant Nos.
11934015, 11975183, 11774397, 11775178, 11775177 and
11947301), Major Basic Research Program of Natural Science of
Shaanxi Province (Grant No. 2017ZDJC-32), Australian
Research Council (Grant No. DP 190101529),  the Strategic Priority Research Program of the Chinese Academy of Sciences (Grant No. XDB33000000), and
Double First-Class University Construction Project of
Northwest University are gratefully acknowledged.
WL Yang acknowledges IoP/CAS
for hospitality during his visit.

\end{document}